\documentstyle[aps,floats,epsfig]{revtex}
\twocolumn

\newcommand{\be}{\begin{equation}}
\newcommand{\ee}{\end{equation}}
\newcommand{\ba}{\begin{eqnarray}}
\newcommand{\ea}{\end{eqnarray}}
\newcommand{\bd}{\begin{displaymath}}
\newcommand{\ed}{\end{displaymath}}
\newcommand{\bi}{\begin{itemize}}
\newcommand{\ei}{\end{itemize}}

\newcommand{\nn}{\nonumber}

\newcommand{\tj}{$t$-$J$\ }

\newcommand{\htc}{high-$T_c$\ }

\newcommand{\fw}{0.8\linewidth}

\newcommand{\krho}{K_{\rm \rho}}
\newcommand{\ie}{i.\,e.\ }

\begin{document}
\wideabs{ 
\title{
  The transition between hole-pairs and four-hole clusters in
  four-leg \tj ladders
  }
\author{Thomas Siller $^{1}$, Matthias Troyer$^{1}$, T.\,M. Rice$^{1}$,
        Steven\,R. White$^{2}$}
\address{
        $1$ Theoretische Physik, 
        Eidgen\"ossische Technische Hochschule, CH-8093 Z\"urich,
        Switzerland \\
        $^2$ Department of Physics and Astronomy, University of California
        Irvine, CA 92697
        }
\date{\today}
\maketitle
\begin{abstract}
Holes weakly doped into a four-leg \tj ladder bind in pairs.
At dopings exceeding a critical doping of $\delta_c\simeq \frac{1}{8}$
four hole clusters are observed to form in DMRG calculations.
The symmetry of the ground state wavefunction does not change
and we are able to reproduce this behavior qualitatively with an
effective bosonic model in which the four-leg ladder is represented
as two coupled two-leg ladders and hole-pairs are mapped on hard 
core bosons moving along and between these ladders.
At lower dopings, $\delta<\delta_c$, a one dimensional 
bosonic representation for hole-pairs works and 
allows us to calculate accurately the Luttinger liquid 
parameter $\krho$, which takes the universal value $\krho=1$
as half-filling is approached.
\end{abstract}
}


\section{Introduction}

Strongly correlated electrons confined to move
on ladders formed from coupled chains have
been an active topic of research in recent 
years. These examples of lightly doped
spin liquids can be very efficiently 
analyzed using the density
matrix renormalization group\cite{prl:srwhite:dmrg,prb:srwhite:dmrg}
(DMRG) method, which allows long ladders to be
accurately simulated. Ladders with an even number 
of legs have spin liquid groundstates at half-filling
and the spin liquid character remains upon doping due
to the binding of holes to form pairs at low concentrations%
\cite{prb:dagottorierascalapino,prb:barnesdagottorieraswanson,%
prl:noackwhitescalapino1,prl:noackwhitescalapino2,prb:gopalanricesigrist,%
prb:tsunetsugutroyerrice,prb:troyertsunetsugurice,science:dagottorice}. 
The underlying physics of the ladder systems has close
similarities to that of a doped resonant valence bond (RVB) 
phase which is widely believed to be realized in the \htc 
superconductors.
The numerically accurate DMRG simulations of the strongly 
correlated \tj model\cite{prb:zhangrice} give a clear picture
of the behavior of hole-pairs but only in qualitative terms
as the density varies.

In this paper our aim is to analyze the DMRG results and
to use them to determine the effective interactions between
hole-pairs which govern their behavior. Since the spin sector
of even leg ladders remains gapped, there are only charge 
degrees of freedom at low energies and these can be represented
in terms of suitable hard core boson models.
These in turn can be analyzed much more easily and completely.
In an earlier paper\cite{cite:thsiller} we described such a 
representation or mapping of the hole-pairs on a two-leg ladder
to hard core bosons on a single chain. This enabled us to
determine the effective repulsive interactions between the hard
core bosons by fitting to the hole density distribution determined
by DMRG for the \tj model.
In the present paper we extend this analysis to four-leg ladders.
The extra transverse degrees of freedom allow the formation of
larger clusters of two hole-pairs (\ie four hole clusters) when
the hole density exceeds a critical concentration $\delta_c\simeq
\frac{1}{8}$. A cluster represents not a bound state of the
hole-pairs but a finite energy resonance which becomes populated
when the chemical potential reaches a certain threshold.

The outline of the paper is as follows. First we 
treat the low concentration region where the density
profiles obtained from DMRG simulations 
show that hole-pairs simply repel each other.
In this region we extend the mapping onto 
the model of hard core bosons that we used earlier for the two-leg 
ladder and obtain a parameterization of the repulsive interactions
and Luttinger liquid exponent $K_\rho$. We compare the evolution
of $K_\rho$ with hole density in the two and four leg ladders. In
both cases $K_\rho\to 1$ as the hole concentration $\delta\to 0$
indicating predominantly superconducting correlations. To treat the
formation of the four hole clusters it is necessary to introduce
additional transverse degrees of freedom. This is done by mapping
the hole-pairs on a four-leg ladder onto a hard core boson model
on a two-leg ladder. Choosing a potential with a repulsive tail but
with a finite energy resonance on a single rung allows us to reproduce
the DMRG results for the four-leg \tj model, both for the kink in
the chemical potential and for the change in the density profile at
the critical density. The paper ends with a concluding section.


\section{Hole Pairs and Clusters}
\label{def:hhcfunction}
\begin{figure}
\begin{center}
\epsfxsize=0.8\linewidth
\epsffile{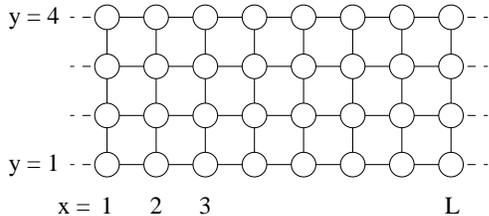}
\end{center}
\caption{
  Indexing convention for a $(L \times 4)$ ladder.\
  }
\label{fig:4legladder}
\end{figure}

We consider the \tj ladder Hamiltonian
\ba
\label{eq:tJham}
H &=& -t \sum_{\langle {\mathbf i},{\mathbf j} \rangle,\sigma} 
{\cal P}(c^\dag_{{\mathbf i}\sigma} c_{{\mathbf j},\sigma} 
+ c^\dag_{{\mathbf j},\sigma}  c_{{\mathbf i},\sigma}){\cal P} \\ \nonumber
&&
+ J \sum_{\langle {\mathbf i},{\mathbf j} \rangle} 
({\mathbf S}_{\mathbf i}{\mathbf S}_{\mathbf j}
  -\frac{1}{4} n_{\mathbf i}n_{\mathbf j}) 
\ea
for a four-leg ladder (4LL) configuration as depicted in Fig.~\ref{fig:4legladder}. 
Here $\sigma=(\uparrow,\downarrow)$ denotes the spin index
and $\langle {\mathbf i}, {\mathbf j}\rangle$ the summation 
over nearest-neighbor sites with ${\mathbf i}=(x,y)$ as site index. 
The operators $c^\dag_{{\mathbf i},\sigma}$ and $c_{{\mathbf i},\sigma}$ 
create or destroy electrons with spin $\sigma$ at position ${\mathbf i}$
respectively.
The corresponding density operator is 
$n_{{\mathbf i},\sigma}=c^\dag_{{\mathbf i},\sigma} c_{{\mathbf i},\sigma}$
and $n_{\mathbf i}=n_{{\mathbf i},\uparrow}+n_{{\mathbf i},\downarrow}$.
The projection operator ${\cal P} \equiv \prod_{\mathbf i}
(1-n_{{\mathbf i},\uparrow}n_{{\mathbf i},\downarrow})$
prohibits double occupancy of a site and ${\mathbf S}_{\mathbf i}$
denotes the spin at site ${\bf i}$. We use $J=0.35\, t$ throughout. 
If not noted otherwise, we use the DMRG method to compute
ground state properties which implies that open boundary conditions
on the ends of the ladder are used.
\begin{figure}
\begin{center}
\epsfxsize=\fw
\epsffile{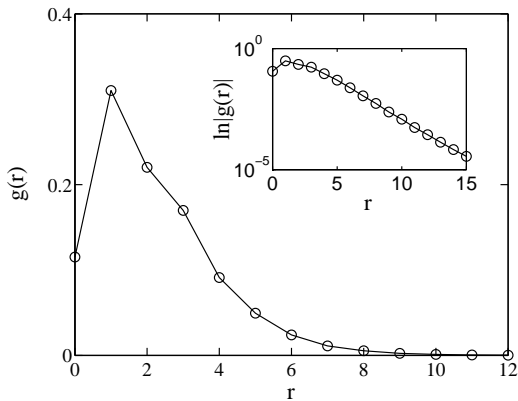}
\end{center}
\caption{
  Hole-hole correlations $g(r)$ as a function of the 
  relative separation $r$ along the legs measured on 
  a ($32 \times 4$) ladder with two holes and 
  $J=0.35\,t$. The inset shows a logarithmic plot of 
  the same data.
  }
\label{fig:hhc}
\end{figure}

We start by examining the internal structure of a hole-pair (HP) on 
a \tj 4LL, by studying the  hole-hole correlation (HHC) function. 
Since the DMRG computations are performed with open boundaries,
special care must be taken in the measurement of the HHC-function,
which we define as
\be
g_{y,y'}(x,x') =
\frac{\langle n_{x,y}^h\,
n_{x'\!,y'}^h\rangle}{\langle n_{x,y}^h\rangle 
\langle n_{x'\!,y'}^h \rangle}\, n^h \ .
\ee
Here $n_{x,y}^h=1-n_{x,y}$ denotes the hole density
operator acting on position $(x,y)$ and $n^h$ is the 
average hole density.
Introducing relative and center of mass coordinates, 
$r=(0,1,\dots)$ and $R_0=(1,3/2,2,\dots)$ in the long direction, 
we define with
\be\label{gr}
g(r,R_0)=\sum_{y,y'} g_{y,y'}(R_0-\frac{r}{2},R_0+\frac{r}{2})
\ee
the HHC between different rungs. 
To obtain a good approximation for the HHC function $g(r)$ 
on an infinite ladder, we measure $g(r,R_0)$ for an ($L\times 4$) ladder 
in the middle of the system, \ie for $R_0\simeq L/2$.
In this way we have measured $g(r)$ on a ($32\times 4$)-ladder 
with open boundaries. The result is plotted in Fig.~\ref{fig:hhc}.
It shows clearly that  in the ground state the two holes are 
bound. The correlation function decreases exponentially as 
can be seen from the logarithmic plot in the inset.
The correlation length is $1.358$ and hence larger than in the 
two-leg ladder (2LL) 
case where we have obtained $1.184$ \cite{cite:thsiller}.
This suggests that the binding energy of a HP is reduced as the 
effective dimension is enhanced. 
With $E(N)$ as the ground state energy with $N$ holes doped 
into the ladder we have computed the binding energy defined 
as
\be
E_b=2\,E(1)-E(0)-E(2)
\ee
for two and 4LL using systems with $40$ and $24$ rungs
respectively. As expected, we have obtained a lower binding energy 
$E_b=0.1062\,t$ for the 4LL than for the 2LL 
case, $E_b=0.1487\,t$.

\begin{figure}
\begin{center}
\epsfxsize=\fw
\epsffile{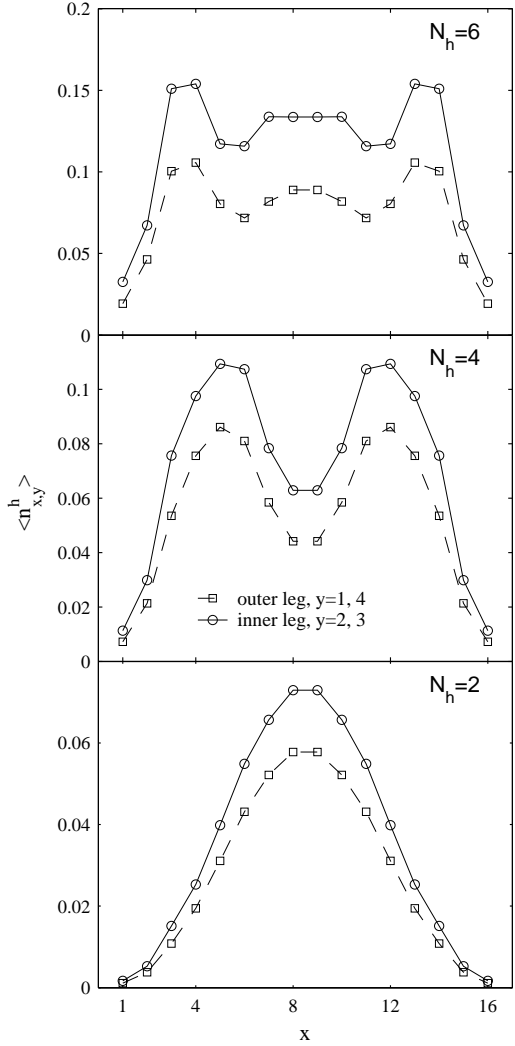}
\end{center}
\caption{
  Hole density profiles $\langle n_{x,y}^h \rangle$ on the inner and outer legs of
  a ($16\times 4$) \tj ladders with   $J=0.35\,t$ for $N_h=2$, $4$ and $6$
  holes doped into the ladder ($\delta < \frac{1}{8}$).
  }
\label{fig:holedoping1}
\end{figure}
\begin{figure}
\begin{center}
\epsfxsize=\fw
\epsffile{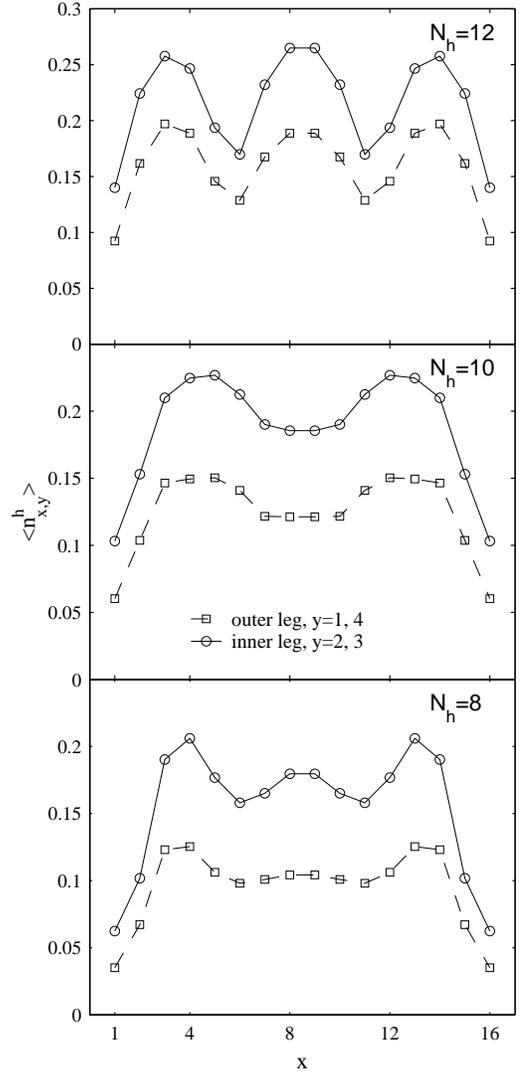}
\end{center}
\caption{Hole density profiles $\langle n_{x,y}^h \rangle$ on the inner and outer legs of
  a ($16\times 4$) \tj ladders with  $J=0.35\,t$ for $N_h=8$, $10$ and $12$
  holes doped into the ladder ($\delta \geq \frac{1}{8}$).
  }
\label{fig:holedoping2}
\end{figure}

%
%
If the HP behave as hard core bosons (HCB) at low
hole doping one should find as many maxima in the
hole density profile of a \tj 4LL with open boundaries 
as the number of HP.
In fact, we observe this only below a critical doping
of $\delta_c \simeq \frac{1}{8}$
in contrast to the results obtained for two-leg 
ladders,\cite{cite:thsiller} where this equality holds
to much higher dopings.
Figures~\ref{fig:holedoping1} and \ref{fig:holedoping2} 
shows the hole density profiles
for the groundstate of a $(16 \times 4)$ ladder for 
$2$ up to $12$ holes. 
The plot with $6$ holes doped into the ladder 
corresponds to a hole density per site 
$\delta < \frac{1}{8}$ and shows three maxima as expected.
For $8$ holes doped into the ladder ($\delta=\frac{1}{8}$)
we could expect to find four maxima, but we observe only
three, suggesting that larger clusters of four holes 
are forming in some way as will be discussed later.


\section{Low density of hole-pairs}

In this section we examine the \tj 4LL at low hole
doping $\delta$, \ie $\delta<\delta_c$. 
\begin{figure}
\begin{center}
\epsfxsize=\linewidth
\epsffile{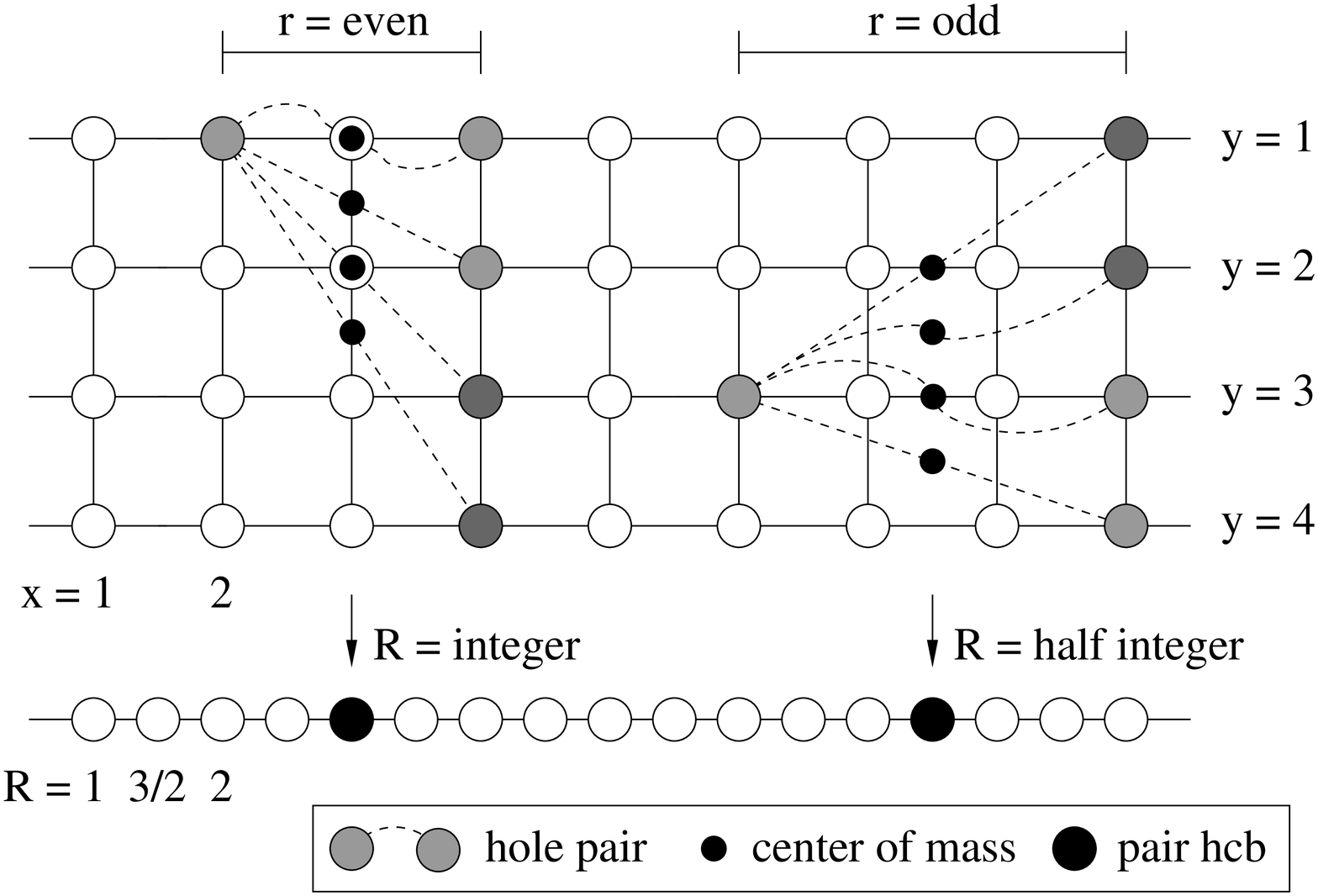}\\
\end{center}
\caption{
  Mapping of hole-pairs from a four-leg ladder to hard core bosons
  on a chain. The center of mass coordinate $R$ of the hole-pair
  determines the hard core boson position.
  }
\label{fig:ladder}
\end{figure}
In order to set up an effective bosonic model we first 
check the symmetry of the ground state at low hole dopings 
using a Lanczos method to get the lowest lying 
energy states with exact diagonalisation for a 
$6 \times 4$ \tj ladder with periodic boundary conditions
along the legs.
For $2$ and $4$ holes doped into the half filled ladder 
the ground state has even parity under reflection along and 
perpendicular to the legs.

Since the spin part of the ground state wavefunction is a 
singlet and thus antisymmetric and the spin excitations are 
gapped,\cite{prb:troyertsunetsugurice} the charge degrees
of freedom can be described by an even parity wavefunction
where HP are considered as effective (hard core) bosons.
At low hole dopings, $\delta<\delta_c$, only one HP 
is found on a rung.
Hence, we can map HP in the $(L \times 4)$ \tj ladder
with periodic boundary conditions in the long direction  
on HCB on a single  closed chain as shown in 
Fig.~\ref{fig:ladder}.
This effective model is completely analogous to the 
model for 2LL introduced in Ref.~\onlinecite{cite:thsiller}.

Since holes on a ladder can pair with more weight on adjacent
rungs, our effective model incorporates the possibility that 
the ``center of mass'' of a HP can lie on a rung or
between two rungs as shown in Fig.~\ref{fig:ladder}.
Note that for even (odd) distance $r$ along the legs between
two holes, the center of mass lies on a rung (between two rungs).
The HHC function $g(r)$ is connected to the probability of 
finding the center of mass of a pair on a rung, $w_{\rm int}$, 
or between two rungs, $w_{\rm half}$, by
\ba\label{wevenodd}
w_{\rm int} &=& \sum_{r_{\rm even}} g(r)\\ 
w_{\rm half}&=& \sum_{r_{\rm odd }} g(r) \nn \ .
\ea

The same occupation probabilities can be obtained with
the one boson Hamiltonian
\ba\label{eq:model2}
H_{\rm B}&=&-t^* \!\!\!\!\! \sum_{R=\frac{1}{2},1,\, \dots}^{L} 
\!\!\!(B_{R}^\dag B_{R+\frac{1}{2}} 
+ B_{R+\frac{1}{2}}^\dag B_{R}) + \epsilon \!\!\! 
\sum_{R=\frac{1}{2},\frac{3}{2},\, \dots}^{L} \!\!\! N_{R}
\ea
for a boson $B_R^\dag$ which moves on a closed chain with length $L$, 
\ie $2\,L$ sites, under the action of an alternating on-site potential
which is $0$ or $\epsilon$ on integer or half integer sites respectively.
Here $N_R=B^{\dag}_R B_R$ and $R=(1/2,1,\dots,L)$.

Figure~\ref{fig:ladder} shows the mapping of HP from the ladder to
effective bosons on a single chain. Note that the center of mass
coordinate $R$ of the pair determines the position of the effective
boson.
Here the ratio of the probabilities $w_{\rm half}/w_{\rm int}$
to find the boson on a site with half integer or integer $R$
depends only on $\epsilon/t^*$ and can easily be obtained as
\be
\frac{w_{\rm half}}{w_{\rm int}} = 
\frac{\epsilon^2+(4t^*)^2-\epsilon \sqrt{\epsilon^2+(4t^*)^2}}
{\epsilon^2+(4t^*)^2+\epsilon \sqrt{\epsilon^2+(4t^*)^2}} \ .
\ee \\
From this equation and Eq.~(\ref{wevenodd}) we obtain for $\epsilon$ 
\be\label{epsilon}
\epsilon=
4\,t^* \sinh{\left( \frac{1}{2}\,\ln{\biggl|{\frac{\sum_{r_{\rm even}}
g(r)}{\sum_{r_{\rm odd}} g(r)}}\biggr|}\right)} \ .
\ee
With $g(r)=g(r,R_0)$ obtained as explained in Sec.~\ref{def:hhcfunction}
for $R_0\simeq L/2$ from a ($32\times 4$) \tj ladder 
we get $w_{\rm half}/w_{\rm int} = 1.18$ and $\epsilon=-0.33\,t^*$.
This result shows that the HP is mainly
centered between two rungs.

Once the boson density $\langle N_R \rangle$ 
for the model (\ref{eq:model2}) has been computed,
we obtain the hole rung density $\langle n_x^h 
\rangle = \sum_{y=1}^4 \langle n^h_{x,y}\rangle$ 
on rung $x$ by the convolution
\ba\label{convolution}
\langle n_x^h \rangle&=&
\frac{1}{2} \frac{\sum_{r_{\rm even}} g(r) (\langle N_{x-\frac{r}{2}}\rangle
 + \langle N_{x+\frac{r}{2}}\rangle)} {\sum_{r_{\rm even}} g(r)}\\
&&+
\frac{1}{2} \frac{\sum_{r_{\rm odd}} g(r) (\langle N_{x-\frac{r}{2}}\rangle
 + \langle N_{x+\frac{r}{2}}\rangle)} {\sum_{r_{\rm odd}} g(r)} \ .\nn
\ea

Next we generalize the one boson model to a finite density including the
interactions between the HCB and also the effect of open boundary 
conditions and write the effective Hamiltonian for the HCB as
\ba
\label{modelHamiltonian}
H_{\rm eff} &=& H_{\rm B} +
V_{\rm int} + V_{\rm b}
\ea
The potential $V_{\rm int}$ gives the interaction between HCB, \ie 
HP in the \tj model. 
Since the simulations are for finite systems, we have to take into
account the interaction of the HP with the boundaries.
The potential  $V_{\rm b}$ has been introduced to describe
this effect.


\subsection{Computing $V_{\rm b}$ and $V_{\rm int}$}

Our procedure is to compute the HCB density $\langle N_R \rangle$, 
convolute it with $g(r)$ according to Eq.~(\ref{convolution}) and 
then to compare it with the hole rung density $\langle n_x^h \rangle$ 
of the corresponding \tj system.
In this way we obtain $V_{\rm b}$ and $V_{\rm int}$ by 
fitting the density profile of the effective model to
that of the \tj model.

However there is a problem in the density profiles of the \tj 
ladders. 
The open boundary conditions induce density oscillations 
at the boundaries which cannot be neglected as in the case
of the 2LL.\cite{cite:thsiller} Due to the higher number
of degrees of freedom the HP in the 4LL gets more distorted,
when it approaches the open ends of the system.
To circumvent this problem we use smooth bounday conditions 
(SBC) as proposed in Ref.~\onlinecite{cite:sbc1} for both, 
the \tj and the effective model.
The SBC introduce smoothly decreasing energy parameters 
into the Hamiltonian as the open ends of the ladder are
approached.
We use SBC only in the long direction which extend
into the \tj ladder until the fifth rung from the 
open ends.
We choose the same smoothing function as used in 
Ref.~\onlinecite{cite:sbc2} for the Hubbard model. 
Details are explained in Appendix \ref{app:sbc}.

\begin{figure}
\begin{center}
\epsfxsize=\fw
\epsffile{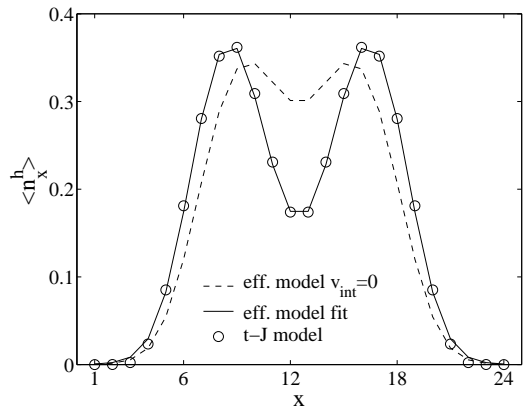}
\end{center}
\caption{
  Hole rung density $\langle n_x^h \rangle$ for $J=0.35\,t$ calculated
  for four holes on a ($24\times 4$) \tj ladder computed directly and with the
  effective model using smooth boundary conditions. For the effective model
  the data for $v_{\rm int}=0$ and the best fit is shown.
  }
\label{fig:intfit}
\end{figure}

We obtain $V_{\rm b}$ by considering one HP
in the \tj model and choose an exponentially decreasing 
form for the potential term $V_{\rm b}$
\ba\label{vb}
V_{\rm b}&=&v_{b} \sum_{R} N_R \, \left(e^{-\frac{R-1}{\xi_{b}}}+
e^{-\frac{L-R}{\xi_{b}}}\right)
\ea
with the two parameters $v_{\rm b}$ and $\xi_{\rm b}$.
\begin{figure}
\begin{center}
\epsfxsize=\fw
\epsffile{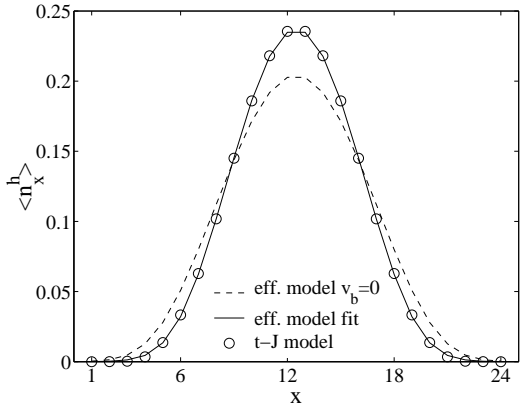}
\end{center}
\caption{
  Hole rung density $\langle n_x^h \rangle$ for two holes on a ($24\times 4$)
  \tj ladder with $J=0.35\,t$ computed directly and with the effective model
  using smooth boundary conditions.
  For the effective model the data for $v_b=0$ and the best fit is shown.
 }
\label{fig:bpfit}
\end{figure}
\begin{table}
\begin{center}
\caption{
  Parameters for the boundary potential $V_{\rm b}$ given in
  Eq.~(\ref{vb}) obtained from the density profile 
  of a ($24 \times 4 $) \tj ladder for  $J = 0.35\, t$.
}\label{tab:bpp}
\end{center}
\begin{tabular}{r r r r}
& $v_{\rm b}/t^*$ & $\xi_{\rm b}$ & \\
\hline
& $148 \times 10^3$ & $0.2980$ & \\
\end{tabular}
\end{table}
Figure~\ref{fig:intfit} shows the results of the fit
and the optimal parameter values are displayed in Table~\ref{tab:bpp}.
Note that the rather large value of $v_{\rm b}$ is a consequence
of the smoothing function, which also decreases the energy 
parameters of $V_{\rm b}$, \ie $v_b$, near the boundaries.

To obtain the interaction potential $V_{\rm int}$ we 
proceed in the same way as for $V_{\rm b}$ and choose
a hard core form
\ba\label{eq:vint}
V_{\rm int} &=& \sum_R \sum_{R'>R} v_{\rm int}(|R-R'|)\, N_R N_{R'} \\
\nonumber
v_{\rm int}(\Delta R) &=& \left\{ 
\begin{array}{ll}
\infty & \textrm{$\Delta R<\Delta R_{\rm min}$} \\
v_1 & \textrm{$\Delta R=\Delta R_{\rm min}$} \\
v\, e^{-(R-R_{\rm min}-\frac{1}{2})/\xi} & \textrm{$\Delta R > \Delta R_{\rm min}$}
\end{array}\right.
\ \ .
\ea
We consider a ($24\times 4$) \tj ladder with
four holes and the corresponding effective model with two HCB
and use three fit parameters for the interaction potential and 
the additional parameter $\Delta R_{\rm min}$, which broadens 
the hard core of the bosons, to model the interaction between 
the pairs.

\begin{figure}
\begin{center}
\epsfxsize=\fw
\epsffile{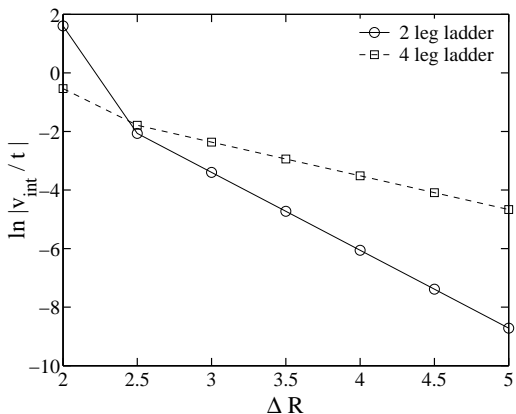}
\end{center}
\caption{%
  Interaction parameter $v_{\rm int}/t$ from Eq.~\ref{eq:vint} 
  obtained for the four-leg ladder in comparison with the 
  result from the two-leg ladder from
  Ref.~\protect\onlinecite{cite:thsiller}.
  }
\label{fig:intpotplot}
\end{figure}

Using the fits to the density profiles as shown
in Fig.~\ref{fig:intfit} the parameter values 
quoted in Table~\ref{tab:fitpar} are obtained.
Figure~\ref{fig:intpotplot} shows the result together with 
that from Ref.~\onlinecite{cite:thsiller} for the two 
leg ladder.

\begin{figure}
\begin{center}
\epsfxsize=\fw
\epsffile{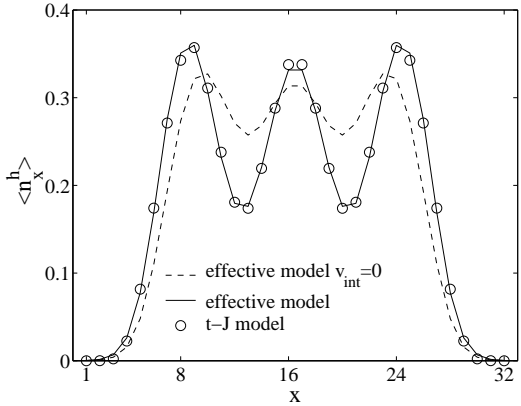}
\end{center}
\caption{
  Hole rung density $\langle n_x^h \rangle$ for $J = 0.35\, t$ for six holes 
  on a ($32\times 4$) \tj ladder using smooth boundary conditions and 
  computed with the interaction potential obtained from fits with two 
  hole-pairs on $24 \times 4$ sites. The data for $V_{\rm int}=0$ for
  the effective model are also shown.
  }
\label{fig:parametertest}
\end{figure}

We have tested the results by comparing the density 
profiles for various numbers of HP and ladder lengths.
We find good agreement between the density profiles
obtained from the \tj model and the effective model 
for hole dopings $\delta\leq 0.08$, as can be seen
from Fig.~\ref{fig:parametertest}.
\begin{table}
\begin{center}
\caption{
  Parameters for the interaction potential $V_{\rm int}$ given in
  Eq.~(\ref{eq:vint}) obtained from the density profile 
  of a ($24 \times 4 $) \tj ladder for  $J = 0.35\, t$.
}\label{tab:fitpar}
\end{center}
\vspace{-4mm}
\begin{tabular}{r r r r r r}
& $v_1/t^*$ & $v/t^*$ & $\xi$ & $\Delta R_{\rm min}$ &\\
\hline
& 0.5848 & 0.2978 & 0.8679 & 2 &\\
\end{tabular}
\end{table}

\subsection{Luttinger liquid parameter $\krho$}

Having determined the effective HCB model that describes the low energy
properties of the HP in the \tj model, we can express the charge
density and the superconducting correlation functions in terms of
Luttinger liquid parameter $K_\rho$ as\cite{luther_peschel}
\ba
\langle N_R N_0 \rangle - \bar N^2&\sim&
{\rm const}\times R^{-2} + {\rm const}\times \cos(\pi (N_b/L) \, R)\, 
R^{-2K_{\rho}}
\label{cdw}
\\
\langle B_R^{\dag} B_0 \rangle &\sim&
{\rm const}\times R^{-{1}/{2K_{\rho}}}\ \ ,
\label{sccorr}
\ea
with $N_b$ as the number of HCB on the chain and $\bar N=N_b/2L$
the mean HCB density per site. 
These relations show that the superconducting
correlations $\langle B_r^{\dag} B_0 \rangle$ are
dominant if $K_{\rho}>\frac{1}{2}$.
For HCB in one dimension, $K_{\rho}$ can
be obtained from the relations \cite{schulz}
\ba
K_{\rho} &=& \pi v_{\rm c} N \biggl( \frac{ \partial^2 E_0}
{\partial N_{\rm b}^2}\biggr)^{-1}\\ 
v_{\rm c} K_{\rho} &=& \frac{\pi}{N}\,\frac{\partial^2 E_0(\Phi)}
{\partial \Phi^2}\bigg|_{\Phi=0} \ \ .\nn
\ea
Here $E_0$ denotes the ground state energy for a closed
ring of length $L$, \ie $N=2\,L$ sites, with $N_{\rm b}$ HCB and
$E_0(\Phi)$ is the ground state energy of the system
penetrated by a magnetic flux $\Phi$ which modifies the
hopping by the usual Peierls phase factor, $ t^* \mapsto t^*
\exp(\pm\, i \Phi/N)$. From these two equations the charge
velocity $v_{\rm c}$ can be eliminated.
We used exact diagonalization for HCB chains with lengths 
between $32$ and $220$ and with $N_{\rm b}=2$.

\begin{figure}
\begin{center}
\epsfxsize=\fw
\epsffile{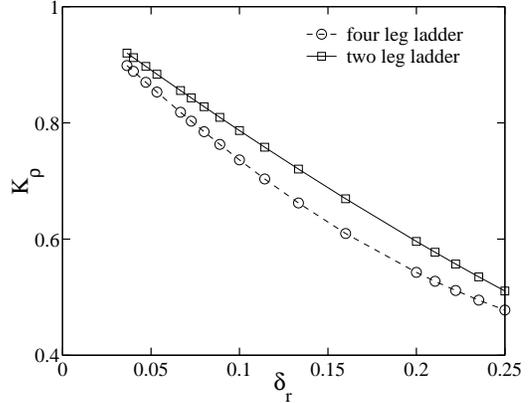}
\end{center}
\caption{
  Luttinger liquid parameter $K_{\rho}$ as a function of hole doping for the 
  four-leg \tj ladder in comparison with the results for the two-leg ladder 
  from Ref.~\protect\onlinecite{cite:thsiller}. The hole doping is given in holes 
  per rung, $\delta_r=N_l\,\delta$, with $N_l$ as the number of legs.
  }
\label{fig:krho}
\end{figure}

The Luttinger liquid parameter $K_{\rho}$ for the interaction
potential given by the parameters in Table~\ref{tab:fitpar}
approaches the universal value $K_{\rho}=1$ at half filling. 
The results are shown in Fig.~\ref{fig:krho} where $K_{\rho}$ is
plotted as a function of the hole doping per rung in the
corresponding \tj ladder, $\delta_r = 4\,\delta = 2\,N_b/N$.
For $N_{\rm b}/N \to 0$, corresponding to a very dilute HCB
gas, we have $K_{\rho}=1+O(N_{\rm b}/N)$ consistent with
Ref.~\onlinecite{schulz_99}. Up to $\delta \simeq 0.06$
($\delta_r\simeq0.23$) the superconducting correlations are
dominant, since $K_{\rho}> \frac{1}{2}$.
 
We have tested numerically the influence of the various parameters 
in Eq.~\ref{eq:vint} on $K_\rho$ and $v_c$. A longer ranged interaction
leads to a lower $K_\rho$ and a larger $v_c$ leaving the product 
$K_\rho\,v_c$ nearly unchanged.
As Fig.~\ref{fig:intpotplot} shows, the interaction between HP 
on 4LL is longer ranged than on 2LL.
Accordingly, the Luttinger liquid parameter for the 4LL
is lower in the investigated density range.
As can be seen from Figure~\ref{fig:krho} the long range charge 
density wave correlations tend to overcome the superconducting
correlations already at lower hole dopings with increasing 
dimensionality, \ie when going from 2LL to 4LL.


\section{The formation of four hole clusters}

In this section we examine the four-hole clusters (FHC) which form on a
\tj 4LL at hole dopings higher than 
a critical doping $\delta_c\simeq \frac{1}{8}$.
\begin{figure}
\begin{center}
\epsfxsize=\linewidth
\epsffile{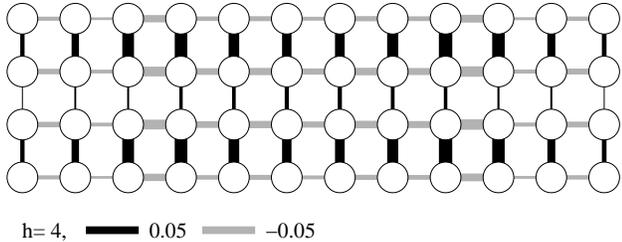}
\end{center}
\caption{
  Real space off diagonal singlet pairing expectation values $s^h_{{\mathbf i},{\mathbf j}}$ 
  between nearest neighbor sites on a ($12 \times 4$) \tj ladder for $h=4$.
  The thickness of the lines gives the magnitude.
  Black and shaded lines denote positive and negative values respectively.
  }
\label{fig:pcorr}
\end{figure}

First we clarify the question whether a change in the
reflection symmetry occurs for the ground state wave
function with increasing doping. We consider the real
space off-diagonal singlet pairing expectation value
(SPE) defined as
\ba
s^h_{{\mathbf i},{\mathbf j}} &=& \langle h-2|\, c_{{\mathbf i},\uparrow}^\dagger 
c_{{\mathbf j},\downarrow}^\dagger - c_{{\mathbf i},\downarrow}^\dagger 
c_{{\mathbf j},\uparrow}^\dagger  \,   |h\rangle
\ea
for nearest neighbor sites $\langle {\mathbf i}, {\mathbf j} \rangle $.
Here $|h\rangle$ denotes the ground state for $h$ holes doped
into the ladder.
We have calculated the SPE on a $12\times 4$ \tj 
ladder for $h$ up to 12, \ie a hole doping $\delta=\frac{1}{4}$.
Figure~\ref{fig:pcorr} shows the results for $h=4$.
The SPE show a characteristic sign-pattern which 
would change with the reflection symmetry over 
the doping, \ie when going from $h \to h+2$.
We observe no change as $h$ is further increased and 
conclude that the ground state symmetry remains unchanged
in the investigated density range. 
This result suggests that HP enter the same band at
dopings below and above $\delta_c$.
Figure~\ref{fig:pcorr} reveals that at low
hole doping HP enter in states where they
reside predominantly on the outer 2LL since
the magnitudes of the SPE between next 
neighbors on these ladders are largest.
But this does not imply that two holes
sit on the same rung. Instead the HHC
function $g_{y,y'}(x,x')$ shows that two
holes can most probably be found in a 
diagonal configuration on a $2 \times 2$ 
plaquette on one of the two coupled 2LL.
\begin{figure}
\begin{center}
\epsfxsize=\linewidth
\epsffile{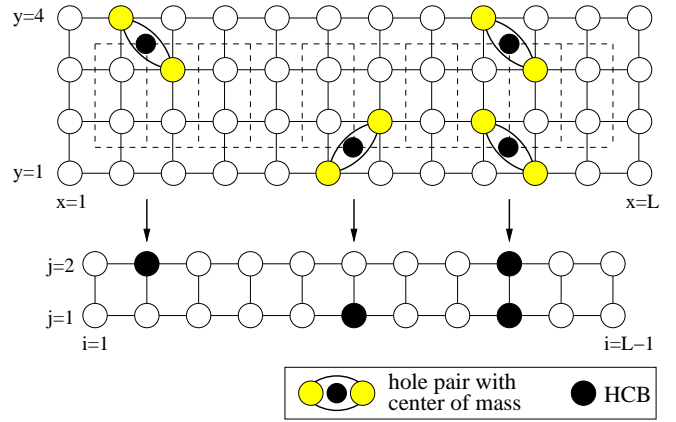}
\end{center}
\caption{
  Mapping of pairs of holes from a four-leg $\tj$ ladder to 
  hard core bosons on a two-leg ladder. In this 
  representation a hole-pair is centered on a $2\times 2$ 
  plaquette on one of the two coupled two-leg ladders.
  }
\label{fig:mapping}
\end{figure}

To set up an effective model which can describe the formation
of the FHC, we again model the charge degrees of freedom of 
HP by effective (hard core) bosons moving in a 
spin liquid but introduce additional transverse degrees
of freedom by considering the \tj 4LL as two
coupled 2LL. This allows us to map HP 
-- represented by their center of mass -- from 
the $(L\times 4)$ \tj ladder on HCB on a 
$((L-1)\times 2)$ ladder as depicted in Fig.~\ref{fig:mapping}.
In this representation a HP is centered on
a $2\times 2$ plaquette on one of the two coupled 
2LL.
The Hamiltonian is given by
\ba\label{eq:effectivemodel2}
H &=&
-t_l \! \sum_{i=1}^{L-2}\sum_{j=1}^2
(B^\dag_{i,j} B_{i+1,j} + B^\dag_{i+1,j} B_{i,j})
\\ &&
 -t_r \! \sum_{i=1}^{L-1}
(B^\dag_{i,1} B_{i,2} + B^\dag_{i,2} B_{i,1})
\nonumber
\\
&&
+ \frac{1}{2}\,\sum_{i,\, j,\, i',\, j'} v(|i-i'|,|j-j'|)\, N_{i, j}\, N_{i',
  j'} \nn
\ea
where $B_{i,j}^\dag$ and $B_{i,j}$ create and destroy a HCB 
on site $(i,j)$ respectively and $N_{i,j}$ is the corresponding density
operator. 
The matrix elements $t_l$ and $t_r$ stand for nearest neighbor hopping 
along the legs and rungs of the 2LL.
The last term describes the interaction between HP and is
determined by the parameter $v(r_x,r_y)$ with $r_x$ and $r_y$ as 
the relative separation of two HCB along and perpendicular to the 
legs respectively.

As hopping matrix element along the legs, $t_l$, we use the value
obtained in Ref.~\onlinecite{cite:thsiller} for 2LL, 
$t_l=0.303 \,t$.
The energy difference of the bonding and antibonding
band is given by $2\,t_r$. 
We have computed this energy difference at wavevector 
$k_x=0$ with exact diagonalisation for two holes on a 
$6 \times 4$ leg \tj ladder using periodic boundary 
conditions in the long direction and obtained 
$t_r=0.092\,t$ so that $t_l/t_r \simeq 3$.

We are able to reproduce qualitatively the formation
of FHC but the formation depends crucially on the form
of the interaction between the HP.
We know from Ref.~\onlinecite{cite:thsiller} that on a 
2LL HP repel each other and hence choose a repulsive
interaction for $r_x > 0$ which tends to zero when
$r_x$ increases.
Further we expect this repulsion to be less for hole
pairs on different legs of the effective 2LL, 
\ie $v(r_x,1)<v(r_x,0)$. At short distance we can 
expect a certain energy lowering due to a lower cost in
magnetic energy in the configuration with four holes on
neighboring sites. Thus we set the constraint that
$v(r_x,1)$ has a local minimum at $r_x=0$. 

\begin{figure}
\begin{center}
\epsfxsize=\fw
\epsffile{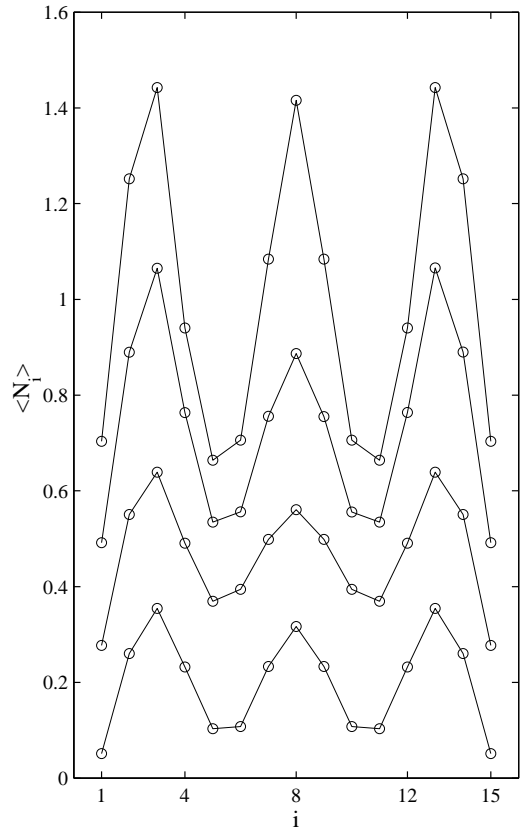}\\
\end{center}
\caption{
  Hard core boson (HCB) density per rung $\langle N_i\rangle$
  obtained from the effective model (\ref{eq:effectivemodel2})
  for a $(15 \times 2)$ ladder with $3$ up to $6$ HCB.
  From below: $3$, $4$, $5$, and $6$ HCB, respectively.
  The plots are shifted by $0.2$ with respect to each other.
  The interaction parameter values from Table~\ref{tab:intparameter} 
  have been used. The data reproduces qualitatively the hole cluster 
  formation shown in Fig.~\ref{fig:holedoping1} and Fig.~\ref{fig:holedoping2}.
  }
\label{fig:hcb15}
\end{figure}

The solution of the two particle problem for the effective
HCB model with this type of interaction shows that in the 
ground state the two HCB are unbound. 
But there are also resonant cluster states which can be occupied
at higher energies. 
Hence, it can be expected that with increasing HCB density
beyond a certain value of the chemical potential a resonant
state is occupied.
This occurs when the expense in kinetic energy due 
to the loss of some degrees of freedom is compensated 
by a smaller interaction energy.
Above the critical doping a HP can cluster with 
another on the same rung but then it tunnels from one suitable
HP to another HP thereby gaining additional kinetic energy. 
\begin{table}
\begin{center}
\caption{
  Interaction parameter values $v(r_x,r_y)$ in units of 
  the hopping matrix element $t$ of the \tj Hamiltonian (\ref{eq:tJham}).
}\label{tab:intparameter}
\end{center}
\begin{tabular}{r r r r r r r r}
  & $r_x$ & $0$ & $1$ & $2$ & $3$ & $\geq 4$&\\
\hline
& $v(r_x,0)/t$ & --- & $100$ & $5$ & $0.1$ & $0$ & \\
& $v(r_x,1)/t$ & $0$   & $0.1$ & $0.5$ & $0.1$ & $0$ &
\end{tabular}
\end{table}
A set of parameter values for $v(r_x,r_y)$ is given in 
Table~\ref{tab:intparameter}. With these we calculated 
the HCB density per rung,
$\langle N_i\rangle = \sum_{j=1}^2 \langle N_{i,j}\rangle$,
for $3$ up to $6$ HCB on a $15\times 2$ ladder shown
in Fig.~\ref{fig:hcb15}.
These results correspond to the hole density profiles
shown in Fig.~\ref{fig:holedoping1} and Fig.~\ref{fig:holedoping2}
for the $16\times 4$ \tj ladder for $3$ up to $6$ HP.
As can be seen, the formation of FHC is reproduced qualitatively.
The number of maxima in the density profiles does not change,
if a fourth HCB (HP in the \tj model) is added and for $6$ HCB
(HP) we observe three well separated maxima, each containing two
HCB (HP).
The density profiles for $5$ HCB and HP look a bit different.
Here the two additional HP cluster with the others but, due to
the repulsive interaction between them, they are pushed towards 
the open ends of the sample in x-direction. We observe the same
effect in the effective model, where the two outer maxima have
higher weights than the maxima in the center.
\begin{figure}
\begin{center}
\epsfxsize=\fw
\epsffile{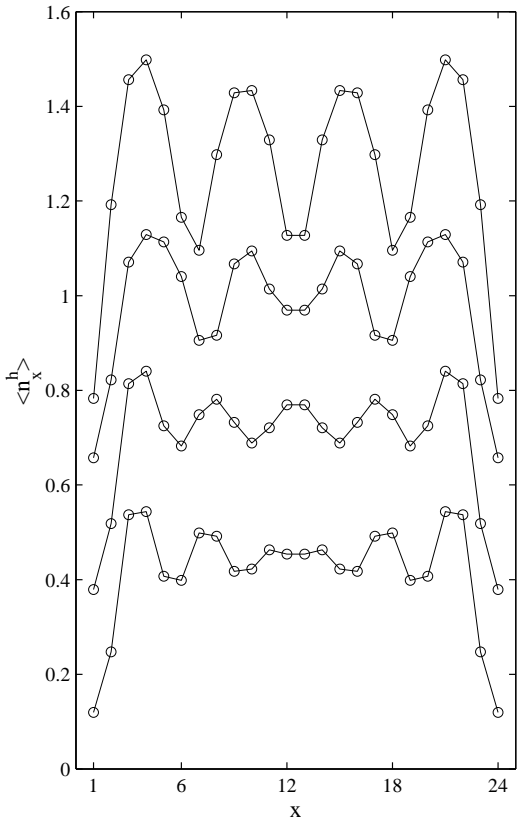}\\
\end{center}
\caption{
  Hole density per rung $\langle n_{x}^h\rangle$
  on a $(24\times 4)$ \tj ladder with $J=0.35\,t$ for $5$ up to $8$ hole
  pairs doped into the ladder. From below: $5$, $6$, $7$, and $8$ hole
  pairs, respectively. The plots are shifted by $0.2$ with respect to
  each other. 
  }
\label{fig:tj24}
\end{figure}
We tested the effective model also for longer systems.
Figure~\ref{fig:tj24} shows the rung density profiles
for $5$ up to $8$ HP on a $(24\times 4)$ \tj ladder.
The FHC start to form when going from $5$ to $6$ HP
doped into the ladder, which corresponds again to a critical
doping $\delta\simeq \frac{1}{8}$. By filling the ladder further
with HP the number of maximum can also decrease. For $7$
HP we observe only four maxima.
The area attached to each maxima suggests that two FHC 
form near the open ends by occupying seven rungs each,
whereas the remaining three HP share the inner ten
rungs, where one pair is loosely bound with the other
two. Finally, for $8$ HP we observe $4$ FHC which occupy
$6$ rungs each. Also this ``contraction'' of HP
can be reproduced with the effective model using the
interaction parameters from Table~\ref{tab:intparameter}.
Figure~\ref{fig:hcb23} displays the corresponding rung
density profiles for $5$ up to $8$ HCB on a $23\times 2$
ladder. They show the same evolution with HCB number
as the hole density profiles in Fig.~\ref{fig:tj24} 
with HP number.
We cannot expect the effective model to reproduce the 
behavior of the \tj model in all details, but in view of the 
simplicity of the model, the results are satisfactory.
After scanning a wide parameter range for $v(r_x,r_y)$ we came
to the conclusion that the formation of FHC can occur when 
the interaction leads to resonant states.
\begin{figure}
\begin{center}
\epsfxsize=\fw
\epsffile{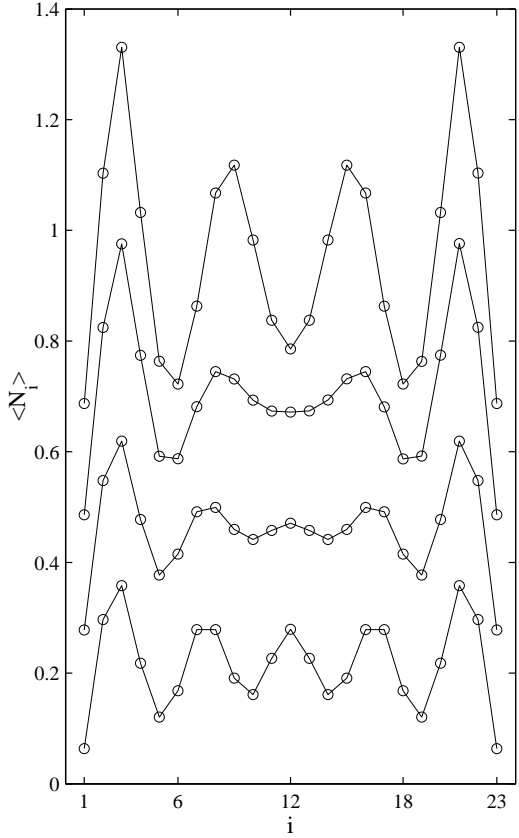}\\
\end{center}
\caption{
  Hard core boson (HCB) density per rung $\langle N_i\rangle$ obtained from
  the effective model (\ref{eq:effectivemodel2}) for a $(23 \times 2)$ ladder
  with $5$ up to $8$ HCB.
  From below: $5$, $6$, $7$, and $8$ HCB, respectively.
  The plots are shifted by $0.2$ with respect to each other.
  The interaction parameter values from Table~\ref{tab:intparameter} have 
  been used. The data reproduces qualitatively the hole cluster formation 
  shown in Fig.~\ref{fig:tj24}.
  }
\label{fig:hcb23}
\end{figure}

Since we observe a qualitative change in the density profiles,
we have computed the chemical potential $\mu(h)$ of the HP
as a function of hole doping
%
\ba
\mu(h)&=&E(h+1)-E(h-1)
\ea
for both models in order to check whether it shows a
singularity at the critical doping. Here, $E(h)$ denotes 
the ground state energy for a system with $h$ holes or 
$h/2$ HCB in the \tj and effective model respectively.
\begin{figure}
\begin{center}
\epsfxsize=\fw
\epsffile{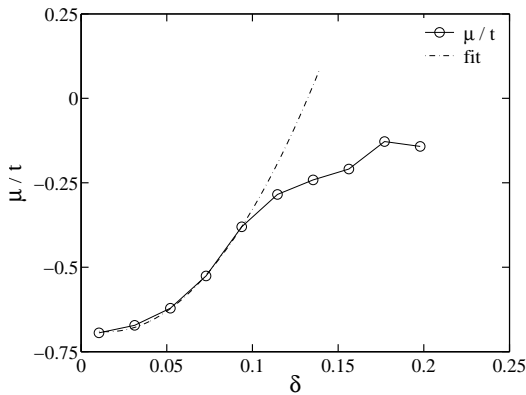}
\end{center}
\caption{
  Chemical potential $\mu/t$ for hard core bosons (hole-pairs) obtained 
  from the bosonic model (\ref{eq:effectivemodel2}) for a 
  ($23\times 2$) ladder using the interaction parameters from
  Table~\ref{tab:intparameter} as function of the hole doping
  $\delta$ in the corresponding \tj model. 
  The dash-dotted line shows a fit to a second order polynomial
  in the region $\delta<\delta_c$.
  }
\label{fig:kink}
\end{figure}
Figure~\ref{fig:kink} displays the chemical potential
for HCB in the bosonic model for a $23\times 2$ ladder
as a function of the hole doping $\delta$ in the
corresponding \tj model. The hole doping is related to
the HCB number $N_b$ by $\delta=N_b/2L$.
The chemical potential shows a deviation from the
quadratic form observed at low densities right at the
position where the formation of the FHC sets in, \ie
when $N_b$ is increased from $5$ to $6$.
The exact position of this kink can be tuned by choosing
appropriate interaction parameter values $v(r_x,r_y)$.
The difference in the chemical potential between the critical
and zero doping, $\Delta \mu \simeq 0.42\,t$, is to a good 
approximation given by the difference between the energies 
of the first resonant state and the ground state, 
$\Delta E=0.43\,t$, obtained by solving the two particle
problem. The critical doping can be shifted to lower values
by making the interaction longer ranged.
\begin{figure}
\begin{center}
\epsfxsize=\fw
\epsffile{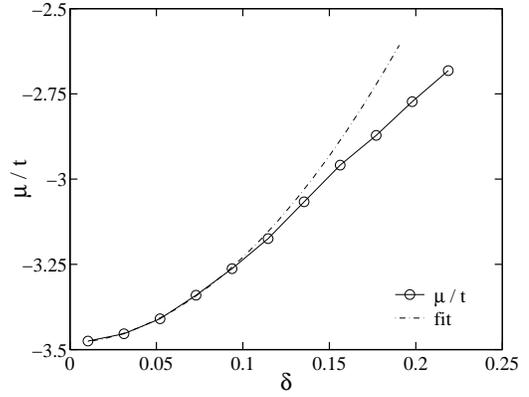}
\end{center}
\caption{
  Chemical potential $\mu/t$ for hole-pairs as a function of hole doping 
  $\delta$ obtained for a ($24\times 4$) ladder with $J=0.35\,t$.
  The dash-dotted line shows a fit to a second order polynomial
  in the region $\delta<\delta_c$.
  }
\label{fig:chempot16}
\end{figure}
Figure~\ref{fig:chempot16} shows the chemical potential for hole
pairs obtained from the corresponding $24 \times 4$ \tj ladder.
The dashed dotted line shows a fit to a second order polynomial 
for $\delta<\delta_c$. As expected, the chemical potential 
$\mu(h)$ deviates from this polynomial for $\delta>\delta_c$.

Finally, we note that while the effective model reproduces the evolution of the
system from HP to FHC nicely, it does not reproduce the density profiles
quantitatively in the FHC density regime as can be seen, for example, by
comparing the upper curves in Fig.~\ref{fig:tj24} and \ref{fig:hcb23}.
Therefore, further refinement of the effective model would be required to
obtain the accuracy necessary to calculate parameters such as $K_\rho$ from
the effective model in this density regime.

\section{Conclusions}

In this work our previous study of the low energy properties of the hole doped
\tj ladders with two legs is extended to the case of four legs. In both cases
density matrix renormalization group (DMRG) results show that the holes bind
in pairs at low densities and a finite
spin gap is preserved. To analyze the DMRG results in more detail we introduce
a hard core boson model on a single chain to describe the low energy degrees
of freedom. The effective interactions between the hard core bosons which
represent hole-pairs, are determined by fitting the density profiles to those
obtained by DMRG methods for the \tj model. This effective repulsive
interaction is longer ranged for the four-leg ladder and this in turn reduces
the value of the Luttinger liquid parameter $K_\rho$. As a consequence the
region of predominantly superconducting correlations is reduced in the wider
ladder. Whereas the hole-pairs in the two-leg ladder simply repel each other,
in the wider four-leg ladder a modification of the hole density profiles
appears beyond a critical hole doping which can be simply interpreted as the
formation of four hole clusters. To reproduce this behavior in the hard core
boson model, it is necessary to introduce an extra transverse degree of
freedom by replacing the single chain with a two-leg ladder. It is also
necessary to modify the interaction potential to incorporate a four hole
cluster as a finite energy resonance. This is achieved by replacing the
monotonically decreasing repulsive interaction on the single chain by one with
a potential minimum at short range. The longer range repulsive tail of the
interaction potential causes the chemical potential to add a hole-pair to rise
as the density is increased. When the chemical potential exceeds a critical
value approximately equal to the resonance energy, the four hole clusters are
formed leading to a kink in the chemical potential as a function of hole
density.

\section{Acknowledgements}
SRW acknowledges the support of the NSF under grant
DMR98-70930. MT was supported by the Swiss National
Science Foundation. The DMRG calculations have been
performed on the SGI Cray SV1 of ETH Zurich.

\appendix
\section{Smooth boundary conditions}
\label{app:sbc}
The SBC introduce smoothly decreasing energy parameters  
into the Hamiltonian as the edges of the lattice are
approached.\cite{cite:sbc1,cite:sbc2}
The result of this operation is that, instead of having a 
sharp and rigid boundary, the boundary extends itself into 
the system and its exact size is not fully determinable.
In this way we talk of the bulk of the system as the region 
where the energy parameters are constant, and of the boundary 
as the region over which the parameters are smoothly turned 
off. 

For the \tj ladders we use SBC only in the long direction and 
denote the width of the boundary with $m$.
\begin{figure}
\begin{center}
\epsfxsize=\fw
\epsffile{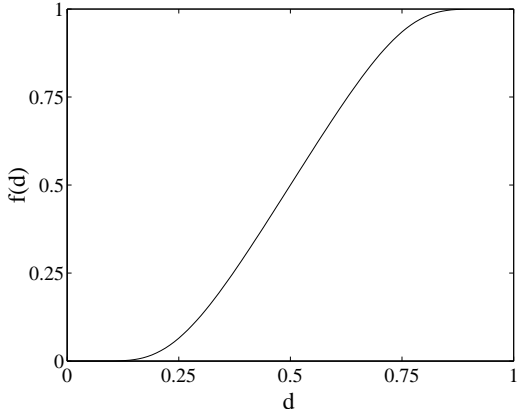}
\end{center}
\caption{
  Smoothing function $f(d)$.
  The parameter $d$ gives the distance 
  to the nearest outermost rung of the system in units of the
  width $m$ of the boundary.
  }
\label{fig:smoothfunc}
\end{figure}
By applying SBC to the Hamiltonians (\ref{eq:tJham}) and (\ref{eq:model2})
we replace the energy parameters $t$, $J$, $t^*$, $\epsilon$, $v_{\rm b}$,
and $v_{\rm int}$ by the site dependent parameters $t_{{\bf i},{\bf j}}$, 
$J_{{\bf i},{\bf j}}$, $t_{R}^*$, $\epsilon_{R}$, $v_{{\rm b},{R}}$,
and $v_{{\rm int},{R,R'}}$ respectively. All these parameters are scaled
according to the smoothing function $f$ defined as
\be
f(d)=\left\{
\begin{array}{ll}
0 & d=0 \\
\frac{1}{2} \left[ 1+\tanh\frac{d-\frac{1}{2}}{d(1-d)} \right] & 0<d<1 \\
1 & d>=1 \ \ \ .
\end{array}
\right.
\ee
and plotted in Fig.\ref{fig:smoothfunc}.
Here $d$ denotes the distance to the nearest outermost rung 
of the system in units of $m$. 
For the \tj model we use 
\ba
d_{{\bf i},{\bf i}'}&=&d_{x,y,x',y'}
=\min(\frac{x+x'}{2}-1, L-\frac{x+x'}{2})/m
\ea
with $L$ as the number of sites in the long direction 
of the \tj ladder 
and scale the energy parameters in such a way that
$t_{{\bf i},{\bf j}}/t=f(d_{{\bf i},{\bf j}})$ and 
$J_{{\bf i},{\bf j}}/J=f(d_{{\bf i},{\bf j}})$, 
where $t$ and $J$ are the bulk values.

For the one dimensional chain with length $N$ 
of the effective model we define
\ba
d_{R,R'}&=&\min(\frac{R+R'}{2}-1, N-\frac{R+R'}{2})/m \\
d_R&=&\min(R-1,N-R)/m \nn
\ea
and scale the energy parameters in the way that
$t_{R}^*/t^*=f(d_{R+\frac{1}{4}})$, $\epsilon_{R}/\epsilon=f(d_R)$, 
$v_{{\rm b},{R}}/v_{\rm b}=f(d_R)$,
and $v_{{\rm int},{R,R'}}/v_{\rm int}=f(d_{R,R'})$. Here again, 
$t^*$, $\epsilon$, $v_{\rm b}$, and $v_{\rm int}$ are the bulk values.

To determine an appropriate value for the width $m$ of the 
boundaries we have considered the curvature of the rung hole density 
profile of a $(24\times 4)$ \tj ladder. For $m<4$ the curvature
is far from being a smooth function of the rung coordinate but for
$m=4$ it is. So we have used $m=4$ for the computations with SBC.

\end{document}